\documentclass[twocolumn, apl, amsmath, amssymb, showpacs, superscriptaddress]{revtex4-1}
\usepackage{epsf}      
\usepackage{graphicx}
\usepackage{color}
\usepackage{soul}
\usepackage{gensymb}
\usepackage{sidecap}
\usepackage{amsmath}
\usepackage{mathtools}

\begin{document}

\title{Effect of the growth orientation on the physical properties of Sr$_2$CoNbO$_6$ thin films}
\author{Ajay Kumar}
\affiliation{Department of Physics, Indian Institute of Technology Delhi, Hauz Khas, New Delhi-110016, India} 
\author{Ramcharan Meena}
\affiliation{Department of Physics, Indian Institute of Technology Delhi, Hauz Khas, New Delhi-110016, India}
\affiliation{ Material Science Division, Inter-University Accelerator Center, Aruna Asaf Ali Road, New Delhi-110067, India}
\author{M. Miryala}
\affiliation{Shibaura Institute of Technology, 3-7-5 Toyosu campus, Koto-ku, Tokyo 135-8548, Japan}
\author{ K. Ueno}
\affiliation{Shibaura Institute of Technology, 3-7-5 Toyosu campus, Koto-ku, Tokyo 135-8548, Japan}
\author{Rajendra S. Dhaka}
\email{rsdhaka@physics.iitd.ac.in}
\affiliation{Department of Physics, Indian Institute of Technology Delhi, Hauz Khas, New Delhi-110016, India}
\date{\today}

\begin{abstract}

We study the effect of the growth orientation on the structural, electronic, and hence transport properties of Sr$_2$CoNbO$_6$ thin films grown on the orthorhombic NGO(100) and cubic MgO(100) substrates. The x-ray diffraction patterns show the growth of the thin film along $a$-axis resulting in the asymmetric ($b\neq c$) in-plane compressive strain in case of NGO(100), whereas along $c$-axis with tensile strain in case of MgO(100) substrate. The temperature dependent resistivity measurements indicate the lower electronic conductivity for the film grown on the NGO(100) substrate, which is found to be correlated with the higher degree of the oxygen deficiencies and hence larger concentration of the insulating Co$^{2+}$ in this sample. Further, the x-ray photoemission spectroscopy measurements show that Sr and Nb are present in the 2+ and 4+ valence state, whereas Co exist in the 2+, 3+ as well as 4+ states, fraction of which was found to vary with the growth orientation. Moreover, the analysis of leakage current using the sum exponent model indicate the presence of two different relaxation mechanisms in these samples. 
\end{abstract}

\maketitle
\section{\noindent ~Introduction}

A stable crystal structure and hence possibility to accommodate the wide range of the elements in the perovskite structure (ABO$_3$; A: rare earth/ alkali earth metals, B: transition metals) give rise to the exotic physical properties such as giant magnetoresistance, spin frustration, multiferroicity, etc.\cite{Perez_PRL_98, Kundu_PRB_05, Shukla_PRB_18, ShuklaPRB23, Giovannetti_PRL_09}, resulting their important technological applications in resistive switching devices, magnetocaloric effect, solid oxide fuel cells, photovolatics, etc. \cite{Sheng_PRB_09, Das_PRB_17, Tao_NM_03, Chakrabartty_NP_18}. Several external perturbations like temperature, mechanical pressure, chemical pressure (doping) have been extensively used to systematically tune these properties \cite{Chen_JACS_14, Saitoh_PRB_97, DuaPRB21}; however, a fine control on the oxygen stoichiometry, which govern most of their physical properties, is still a major challenge in the family of complex oxides \cite{Deb_PRB_06, Sankar_PRB_05}. In this regard, substrate induced strain in the epitaxial thin films has become novel tool to engineer the oxygen concentration, and resulting electronic structure and magnetic properties of these samples \cite{Petrie_AFM_16, Herklotz_JPCM_17, Aschauer_PRB_13, Ohtomo_Nature_04, Fuchs_PRB_07}. Also, the single crystalline oxide substrates with different lattice parameters can give the flexibility to tailor the oxygen content for the desired physical properties \cite{Kumar_JAP_20, Esser_PRB_18}. Moreover, the double perovskite oxides with the general formula A$_2$BB$^\prime$O$_6$ have further attracted the research community, where the degree of rock salt like ordering in the BO$_6$ and B$^\prime$O$_6$ octahedra has been widely used to engineer their physical properties \cite{Niebieskikwiat_PRB_04, Sanchez_PRB_02, Jung_PRB_07}. The recent neutron powder diffraction and the x-ray absorption spectroscopy (XAS) measurements on the La substituted Sr$_{2-x}$La$_x$CoNbO$_6$ samples demonstrate that the degree of octaheral distortion plays a key role in controlling the magnetic and electronic properties of these samples \cite{Kumar_JPCL_22, Kumar_PRB3_22}. Interestingly, the substrate induced strain in the thin film samples can significantly rotate/ tilt the (B/B$^\prime$)O$_6$ octahedra by manipulating the B/B$^\prime$--O bond length and/or  B/B$^\prime$--O--B/B$^\prime$ bond angles \cite{Fuchs_PRB_07, Vailionis_PRB_11, Iliev_PRB_07}. For example, Kleibeuker $et$ $al.$ proposed an interesting approach of growing the ordered thin films from the disordered bulk double perovskite target materials, where the formation of the B-site cages of two different volume has been observed, resulting from the tilting of the two adjacent (B/B$^\prime$)O$_6$ octahedra towards in-plane and out-of plane directions of the (111) oriented substrate, respectively \cite{Kleibeuker_NPGAM_17}. 

More importantly, the epitaxial thin films of the Co-based perovskite oxides are of particular interest due to the various possible valence and spin-states of Co, which can be easily altered by changing the crystal field splitting with the help of  the misfit induced strain in the epitaxial thin films \cite{Fuchs_PRB_07, Galceran_PRB_16, Callori_PRBR_15, ShuklaTSF20}. Chen $et$ $al.$ reported that a mechanical pressure of 40~GPa on SrCo$_{0.5}$Ru$_{0.5}$O$_{3-\delta}$ sample can completely transforms the Co$^{3+}$ ions from high spin (HS) to low spin (LS) state due to reduction in the Co--O bond distance \cite{Chen_JACS_14}. This suggest that the substrate induced strain can be used as an alternative tool to manipulate the electronic structure of such compounds in the epitaxial thin films. For example, even a small degree of the compressive strain in the La$_2$CoMnO$_6$ thin films (grown on LSAT and LaAlO$_3$ substrates) favors the in-plane magnetic anisotropy, whereas a tensile strain (on SrTiO$_3$ substrate) results in the out-of-plane magnetic anisotropy due to difference in the cell parameter in the two cases \cite{Galceran_PRB_16}. In this line, Sr$_2$CoNbO$_6$ due to its moderate charge and ionic radii difference between two B-site atoms (Co$^{3+}$ and Nb$^{5+}$), which is crucial for achieving the B-site ordering \cite{King_JMC_10, Vasala_SSC_15}, results in the intriguing physical properties like colossal dielectric response and complex ac impedance spectroscopy \cite{Xu_JCSJ_16, Wang_AIP_13, Bashir_SSS_11}. Recently, we have extensively studied the magnetic, transport, and electronic properties of La substituted Sr$_{2-x}$La$_x$CoNbO$_6$ bulk samples, which shows the low temperature cluster-glass-like behavior for Sr$_2$CoNbO$_6$ and insulating/semiconducting nature with the electronic activation energy of 0.27~eV \cite{Kumar_PRB2_20, Kumar_PRB1_20}. Also, Wang $et$ $al.$ reported the colossal dielectric properties in Sr$_2$CoNbO$_6$ and its strong correlation with the conductivity of the sample \cite{Wang_AIP_13}. Theoretical calculations by He $et$ $al.$ claim the Sr$_2$CoNbO$_6$ as an indirect band gap semiconductor with a band gap of 2.926~eV \cite{He_JPCM_20}. In order to  tune the electronic band structure in a controlled manner, the epitaxial thin films of Sr$_2$CoNbO$_6$ were grown on different substrates with the varying degree as well as the direction of the substrate induced strain, where the compressive and tensile strains were found to decreases and increases the electronic band gap, respectively \cite{Kumar_JAP_20}. The analysis suggest that the cumulative effect of the substrate induced strain, oxygen non-stoichiometry, and degree of covalency character in the bonding  govern the underlined transport mechanism in the compound \cite{Kumar_JAP_20}. Thus, a precise understanding of its electronic structure is necessary for the device fabrication. However, the effect of growth orientation of the films on the electronic structure has not been studied, which can be useful to disentangle the contribution of various factors on the electronic structure of Sr$_2$CoNbO$_6$. Moreover, an estimation of the steady-state leakage current is also crucial for it's possible use in the energy storage devices, dynamic random access memory (DRAM), transistors etc. \cite{Nagaraj_PRB_99, Zhang_PRA_20, Podgorny_AIP_16}. 

Therefore, we investigate the epitaxial thin films of Sr$_2$CoNbO$_6$ on orthorhombic NGO(100) substrate with the asymmetric in-plane compressive strain and cubic MgO(100) substrate with the symmetric tensile strain. The out-of-plane x-ray diffraction (XRD) measurements reveal the growth of the thin films along $a$ and $c$-axis on the NGO(100) and MgO(100) substrates. A periodic pattern in the surface topography is observed with the average roughness of around 4~nm for both the films using atomic force microscopy. The temperature dependent resistivity show the lower electronic conductivity in case of film grown on the NGO(100) substrate, which is supported by the higher activation energy and lower effective density of states near the Fermi level as compared to that on MgO(100) substrate. The x-ray photoemission measurements clearly show the higher oxygen deficiencies and resulting larger fraction of the Co in 2+ valence state in case of the film on NGO(100) substrate, causing its lower electronic conductivity. Further, the possible mechanisms governing the leakage current in the film on NGO(100) substrate, using the dc step voltage, has been discussed to understand its possible use in the charge storage devices. 

\section{\noindent ~Experimental}

The details about the sample preparation and the physical properties of the bulk Sr$_2$CoNbO$_6$ target sample can be found in Refs. \cite{Kumar_PRB1_20, Kumar_PRB2_20, Kumar_PRB3_22}. The thin films of 70$\pm$5~nm were grown from the bulk target material using the pulsed laser deposition (PLD) technique \cite{ShuklaTSF20} on NdGaO$_3$(NGO)(100)  and MgO(100) substrates. In order to optimize the thickness of the films, we first deposited a film on Si(100) by partially shadowing the substrate to make a sharp step and estimated the film thickness using a stylus profiler across the step. We then used the same deposition parameters for both the films, which were grown at 800\degree C temperature and 10$^{-3}$ mbar oxygen partial pressure using 1.5--2 Jule cm$^{-2}$ laser fluence and 5Hz repetition rate. A post-growth annealing for 15 minutes was performed at 50~mbar oxygen pressure for both the samples at the deposition temperature.  

We carried out the XRD measurements in the Bragg- Brentano geometry using PANalytical X'Pert$^3$ diffractometer. The atomic force microscope (AFM) was used in the non-contact mode to study the surface topography of the films. The low temperature resistivity measurements were performed using  physical property measurement system (PPMS) of Quantum design, USA, with the four probe contact method using 0.1~$\mu$A excitation current. The I-V and leakage current measurements were performed in somewhat unconventional two probe configuration on top of the films using the 2612B source meter and 6517B electrometer from Keithely. The x-ray photomission spectroscopy (XPS) measurements were done with the monochromotic Al-K$\alpha$ (h$\nu=$ 1486.6 eV) x-ray source. Due to the insulting nature of both the samples, a charge neutralizer was used during the measurements. Each spectrum is calibrated {\it w.r.t.} the offset in C 1$s$ peak from 284.6~eV and then fitted with the mixed Lorentzian and Gaussian peak shape after subtracting the inelastic Tougaard background. We use Renishew inVia confocal microscope with 514.5~nm laser source and 1 mW laser power to perform unpolarized Raman spectroscopy measurements. 

 \begin{figure*}
\includegraphics[width=1.0\textwidth]{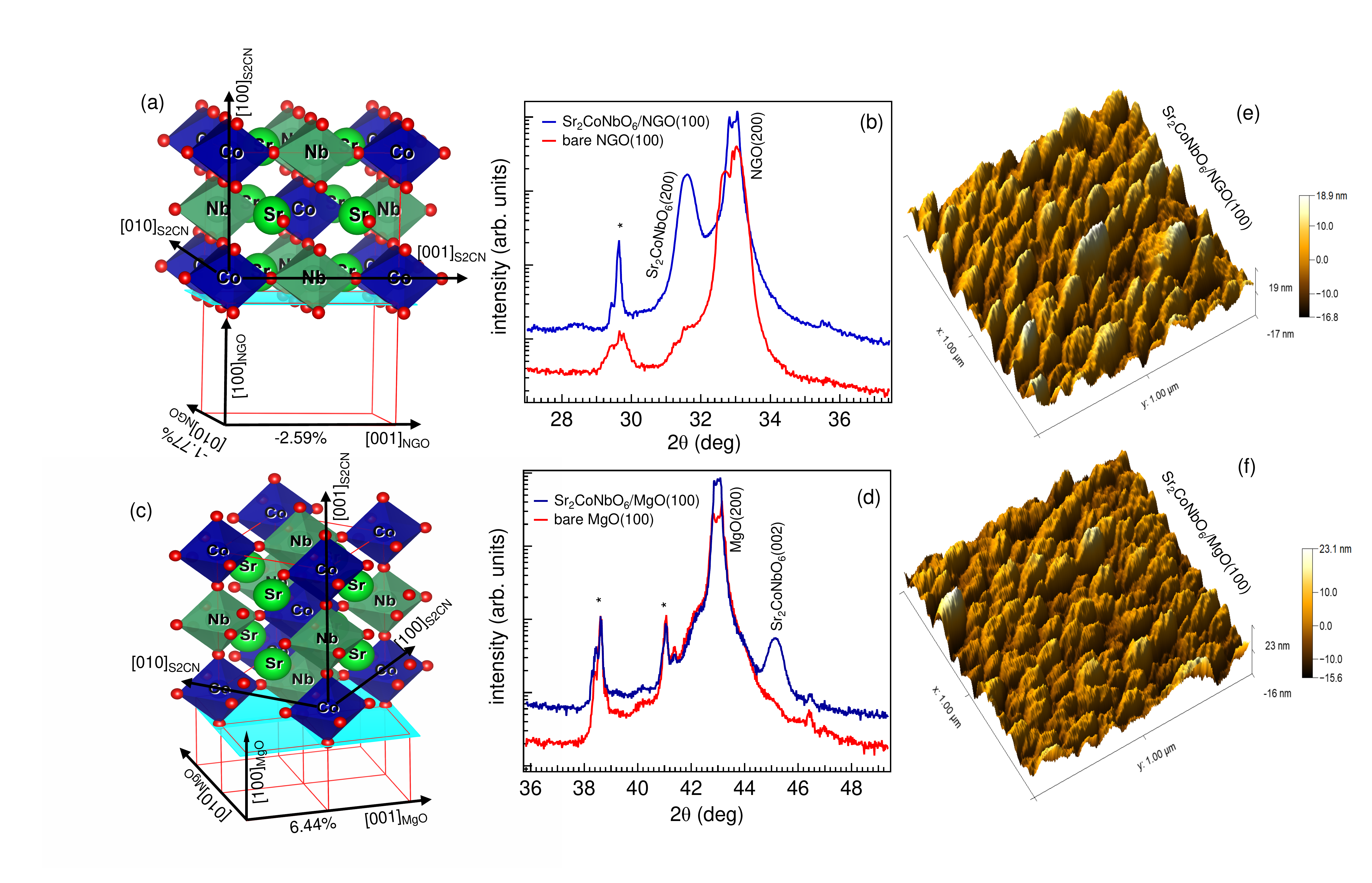} 
\caption {(a) Schematic illustration of the growth orientation of the Sr$_2$CoNbO$_6$ (S2CN) thin films deposited on NGO(100) and (c) MgO(100) substrates. (b) Room temperature $\theta$-2$\theta$ x-ray diffraction pattern of Sr$_2$CoNbO$_6$ thin films grown on NGO [around (200) reflection] and (d) MgO [around (002) reflection] substrates. The asterisk symbols indicate the peaks originating from the substrates. (e) (1x1) $\mu$m AFM images of the same films grown on NGO(100) and (f) MgO(100) substrates.}
\label{Fig1_XRD}
\end{figure*}

\section{\noindent ~Results and discussion}

The orthorhombic crystal structure of NGO substrate with all the three different lattice axes ($a_{\rm NGO}$=5.433~\AA, $b_{\rm NGO}$=5.503~\AA, $c_{\rm NGO}$=7.716~\AA) results in the several possible growth orientations with varying degree of the asymmetric in-plane lattice mismatch on the different surface planes \cite{Huang_PRB_12, Boschker_PRB_09}. In the present case, Sr$_2$CoNbO$_6$ with the tetragonal structure ($a_{\rm S2CN}$=$b_{\rm S2CN}$=5.602~\AA $\space$ and $c_{\rm S2CN} =$7.921~\AA $\space$ \cite{Kumar_PRB1_20}) can be epitaxially grown on the NGO(100) substrate, such that $b$ and $c$-axis lies in-plane of the substrate and $a$-axis in the out-of-plane direction. This produces an asymmetric in-plane lattice misfit [($b_{\rm NGO}$-$b_{\rm S2CN}$)/$b_{\rm S2CN}$] of -1.77\% and -2.59\% along the $b$ and $c$-axis, respectively. The presence of asymmetric in-plane strain in the Sr$_2$CoNbO$_6$ film grown on NGO(100) substrate is schematically illustrated in Fig.~\ref{Fig1_XRD}(a). The longer out-of-plane lattice parameter of Sr$_2$CoNbO$_6$ as compared to NGO(100) substrate results in its corresponding peak at lower 2$\theta$ value in the XRD measurements, as shown in Fig.~\ref{Fig1_XRD}(b). The calculated out-of-plane ($a$-axis) lattice parameter of Sr$_2$CoNbO$_6$ from the XRD data is found to be 5.656~\AA, which is 0.96\% longer than the bulk lattice parameter (5.602~\AA), indicating the significant effect of the in-plane compressive strain on the lattice structure of Sr$_2$CoNbO$_6$, which results in the change in orbital hybridization, and consequently the electronic and transport properties in the thin films, discussed later. Also, the presence of oxygen deficiencies in the perovskite thin films are also widely known to expand the unit cell \cite{Kim_PRB_20, Tyunina_SR_21} and hence a cumulative effect of both substrate induced strain and oxygen vacancies (discussed below) are expected to govern the lattice parameter in the present case. The peak marked by asterisk symbol in Fig.~\ref{Fig1_XRD}(b) results from the  crystal imperfection in the substrate, as evident from the XRD pattern of the bare NGO(100) substrate. 

Further, the epitaxial growth of Sr$_2$CoNbO$_6$ on the cubic MgO(100) substrate ($a_{\rm MgO}$=4.216~\AA) is illustrated in Fig.~\ref{Fig1_XRD}(c), where $a$ and $b$-axis lie in-plane of the substrate and $c$-axis along the out-of-plane direction. The in-plane lattice parameters of Sr$_2$CoNbO$_6$ in the pseudocubic representation are $a/\sqrt{2}$=3.961~\AA, which results in a large lattice misfit of 6.44\% with the MgO(100) substrate producing tensile strain. This relatively large in-plane lattice misfit may also lead in the gradual relaxation in the film as we move away from the interface and reciprocal space mapping (RSM) measurements can be further helpful to directly probe the strain state in these films \cite{Nichols_APL_13}. The presence of the film peak at the higher 2$\theta$ value in the XRD pattern as compared to the substrate, as shown in Fig.~\ref{Fig1_XRD}(d), indicate the smaller out-of-plane pseudocubic lattice parameter (in the present case, $c_{\rm pc}=c/2$, where subscript pc represent the pseudocubic). It is important to note here that the out-of-plane lattice parameter of Sr$_2$CoNbO$_6$ film grown on MgO(100) substrate found to be 4.006~\AA, which is 3.8\% larger than that of bulk sample in spite of the tensile in-plane strain in the film. This indicate the dominant role of oxygen vacancies in the anisotropic expansion of the unit cell of film grown on the MgO(100) substrate \cite{Tyunina_SR_21}. Further, the AFM images for both the samples are recorded to study their surface topography, as shows in Figs.~\ref{Fig1_XRD}(e) and (f) for the films grown on NGO(100) and MgO(100) substrates, respectively. The periodic patterns are observed on the surface of both the samples in (1x1)$\mu$m scan areas. However, these features are more clearly visible in case of the film grown on NGO(100), possibly due to the  relatively smaller lattice misfit of the film with the substrate as compared to the MgO(100) substrate \cite{Kumar_JAP_20}. The calculated average roughness of the films are 4.3~nm in case of NGO(100) and 3.9~nm in case of MgO(100) substrate, which is slightly higher considering the epitaxial growth. The post-growth annealing of the films in the higher oxygen partial pressure may be the possible reason for this large surface roughness.  

\begin{figure}[h]  
\includegraphics[width=0.47\textwidth]{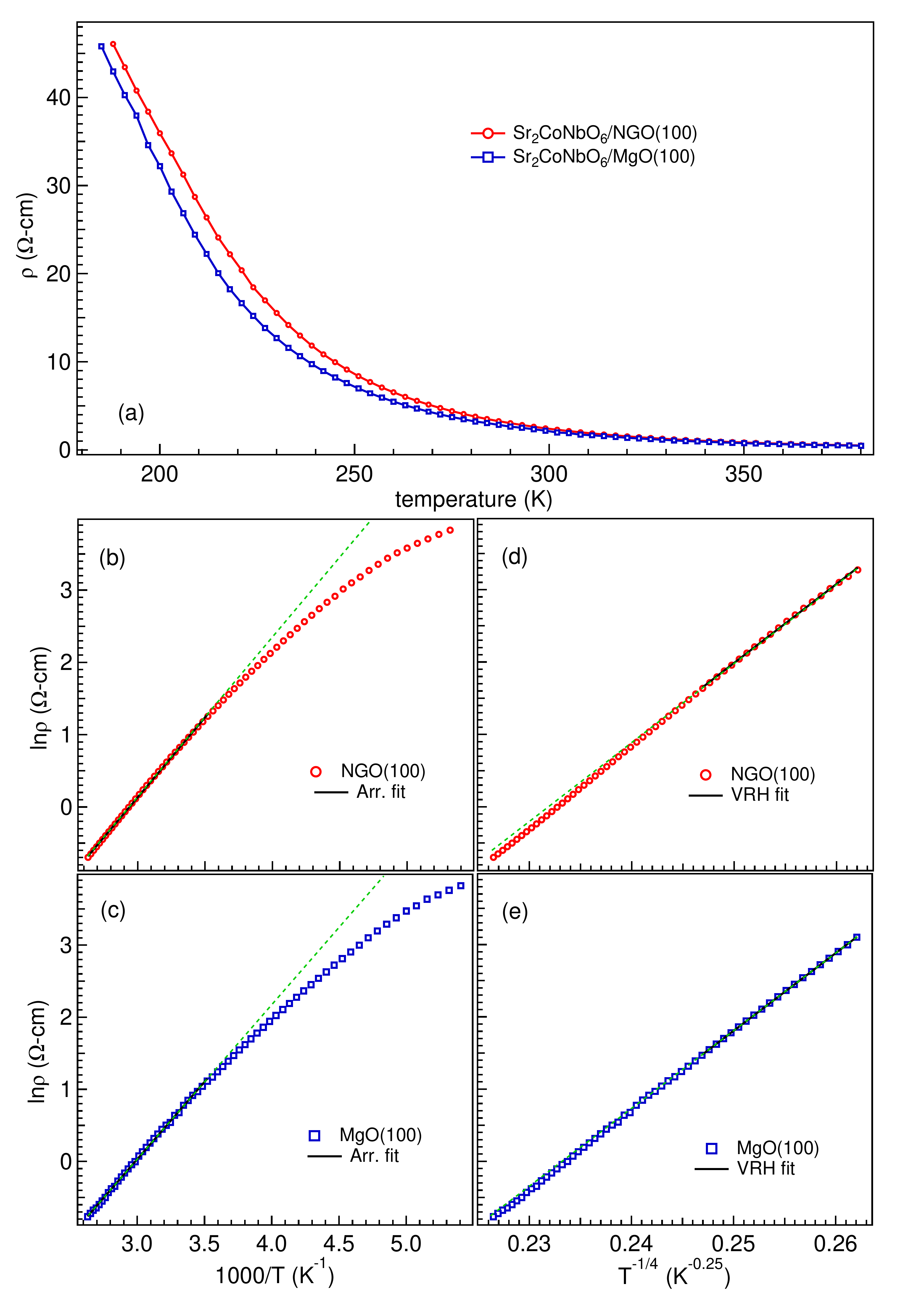}
\caption {(a) The temperature dependent resistivity of Sr$_2$CoNbO$_6$ films grown on NGO(100) and MgO(100) substrates. (b, c) The Arrhenius fitting of the high temperature resistivity data, and (g, h) the VRH fitting in the low temperature regime, for the films grown on NGO(100) and MgO(100) substrates, respectively. Here, the green dashed lines represent the deviation from the Arrhenius and VRH models in the lower and higher temperature regimes, respectively.} 
\label{Fig2_rho}
\end{figure}

It has been well established that the degree and direction of the lattice strain is strongly related to the presence of oxygen non-stoichiometry in the epitaxial thin films of the oxide materials due to change in the molar volume \cite{Petrie_AFM_16, Herklotz_JPCM_17, Aschauer_PRB_13}. Further, the presence of compressive and tensile substrate induced in-plane lattice strain are known to respectively strengthen and weaken the $p$-$d$ orbital hybridization between the oxygen and transition metal atoms, resulting in the enhancement and reduction of the electronic conductivity in the two cases, respectively \cite{Huang_APL_19, Kumar_JAP_20}. Thus, in order to probe the effect of growth orientation and asymmetric in-plane strain on the electronic band structure of Sr$_2$CoNbO$_6$, we record the temperature dependent resistivity of both the films from 185--380~K, as shown in Fig.~\ref{Fig2_rho}(a). We observe semiconducting/ insulating behavior in the measured temperature range for both the films. It is interesting to note that the film grown on the MgO(100) substrate shows the higher electronic conductivity as compared to that on the NGO(100) substrate, despite the compressive in-plane strain in the latter. This suggest that the effect of growth orientation of the films on their electronic structure is more prominent as compared to strain induced change in the metal-ligand orbital hybridization. 

 In order to understand the electronic transport mechanism in the films, the resistivity curves were modeled with the Arrhenius equation, described as
\begin{eqnarray}
 \rho(T)= \rho(0)exp(E_a/k_BT), 
\end{eqnarray}
where E$_a$ is the activation energy require for the conduction of the charge carriers and $\rho$(0) is the pre-exponential factor. The linear fitting of the ln($\rho$) versus 1/T plots from 284--380K gives activation energies of 196~meV and 190~meV for the films deposited on NGO(100) and MgO(100) substrates, as shown in Figs.~\ref{Fig2_rho}(b) and (c), respectively. This indicate the higher conductivity of the film grown on the MgO(100) substrate, as also evident from the Fig.~ \ref{Fig2_rho}(a). However, the conduction behavior of the films deviates from the Arrhenius model at the low temperatures, as shown by the green dashed lines in Figs.~\ref{Fig2_rho}(b) and (c). We find that the conduction mechanism of the films in the low temperature region follow the variable range hopping (VRH) model of the localized charge carriers, given as
\begin{eqnarray}
 \rho(T)= \rho(0)exp(T_0/T)^{1/4}, 
\end{eqnarray}
where T$_0$ is the characteristic temperature defined as, T$_0$ = 18/k$_B$N(E$_F)L^3$, where N(E$_F$) is the localized density of states (DOS) near the Fermi level and $L$ is the hopping length of the charge carriers. The slope of the ln($\rho$) versus T$^{-1/4}$ curves in the low temperature region gives the value of T$_0$, as shown by black solid lines in Figs.~\ref{Fig2_rho}(d) and (e) for both the films grown on the NGO(100) and MgO(100) substrates, respectively. The estimated values of the effective DOS near the Fermi level are 17.9$\times$10$^{19}$ and 18.9$\times$10$^{19}$ eV$^{-1}$cm$^{-3}$ for the films on NGO(100) and MgO(100) substrates, respectively, where we assume Co--O$\approx$2\AA$\space$ as the localization length of the charge carriers. The low DOS value near the Fermi level and higher activation energy required by the charge carriers to take part in the conduction mechanism in case of the film deposited on the NGO(100) substrate results in its lower electronic conductivity as compared to the film deposited on the MgO(100) substrate. A similar reduction in the electronic conductivity of an-isotropically strained La$_{0.67}$Ca$_{0.33}$MnO$_3$ films grown on NGO(100) and NGO(001) substrates has been attributed to the rotation/tilting and distortion in the MnO$_6$ octehedra, as compared to the unstrained (bulk like) film on NGO(110) substrate \cite{Huang_PRB_12}. A higher degree of distortion in the (Co/Nb)O$_6$ octahedra in case of NGO(100) substrate due to the anisotropic in-plane strain can also be the possible reason for this observed reduction in the electronic conduction in the film grown on NGO(100) substrate. 

\begin{figure} [h] 
\includegraphics[width=0.48\textwidth]{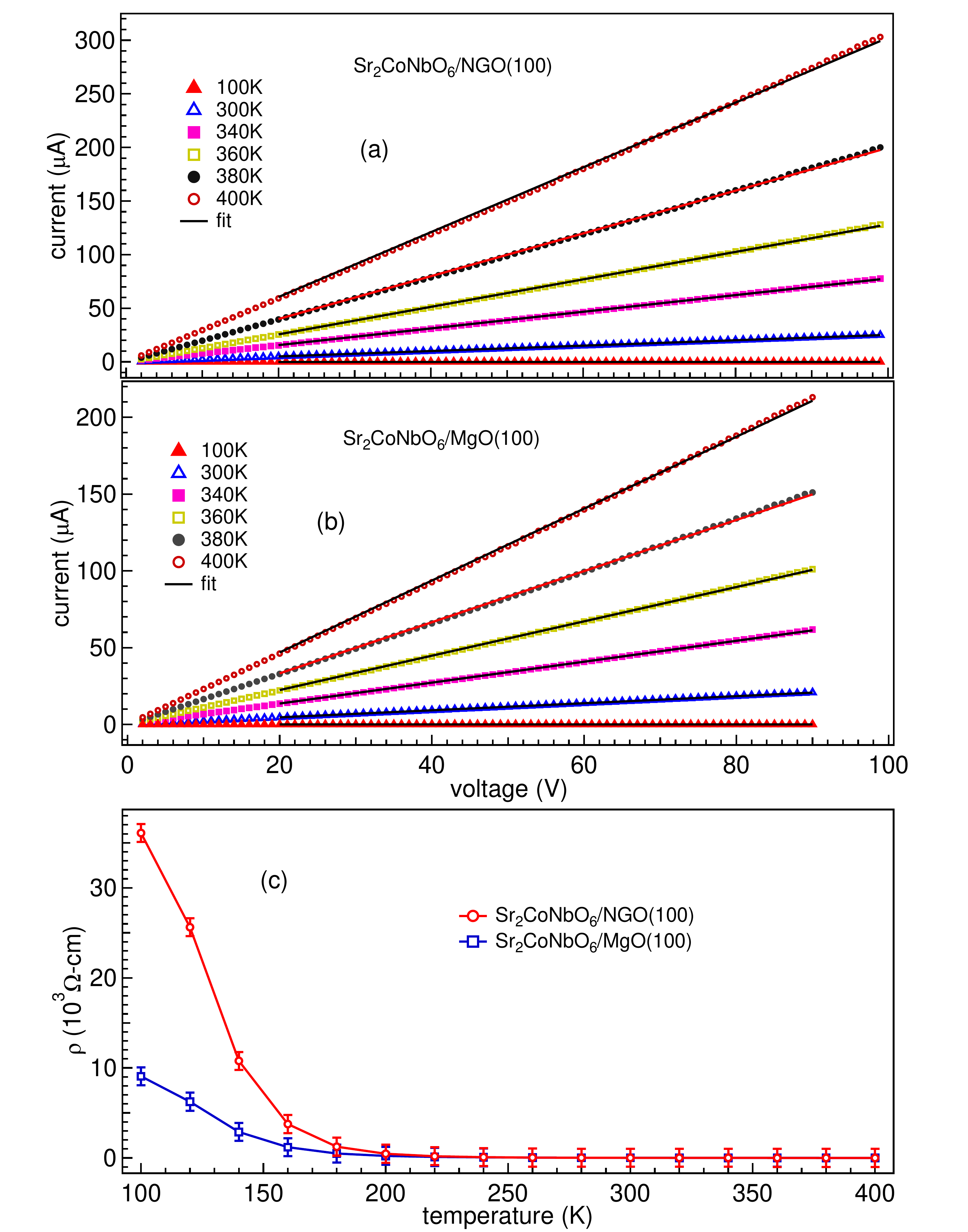}
\caption {The I-V data of the films grown on (a) NGO(100) and (b) MgO(100) substrates at the selected temperatures. (c) The temperature dependent resistivity curves extracted from the slope of I-V curves in the higher voltage regime, as shown by the solid black lines in (a) and (b).} 
\label{Fig3_IV}
\end{figure}

In order to further quantify this effect, the temperature dependent current (I)-voltage (V) measurements are performed and the data are shown in Figs.~\ref{Fig3_IV}(a, b) for the films grown on NGO(100)and MgO(100) substrates, respectively. The I-V curves show the linear behavior, particularly in the high voltage region and the slope of linear fit [shown by the solid lines in Figs.~\ref{Fig3_IV}(a, b)] is used to calculate the conductance and hence resistivity of the samples as a function of temperature, as presented in Fig.~\ref{Fig3_IV}(c). The observed higher resistivity of the film grown on the NGO(100) substrate as compared to that on MgO(100) in I-V data is consistent with the above $\rho$-T data, measured at the fixed excitation current. 

\begin{figure*}
\includegraphics[width=1.02\textwidth]{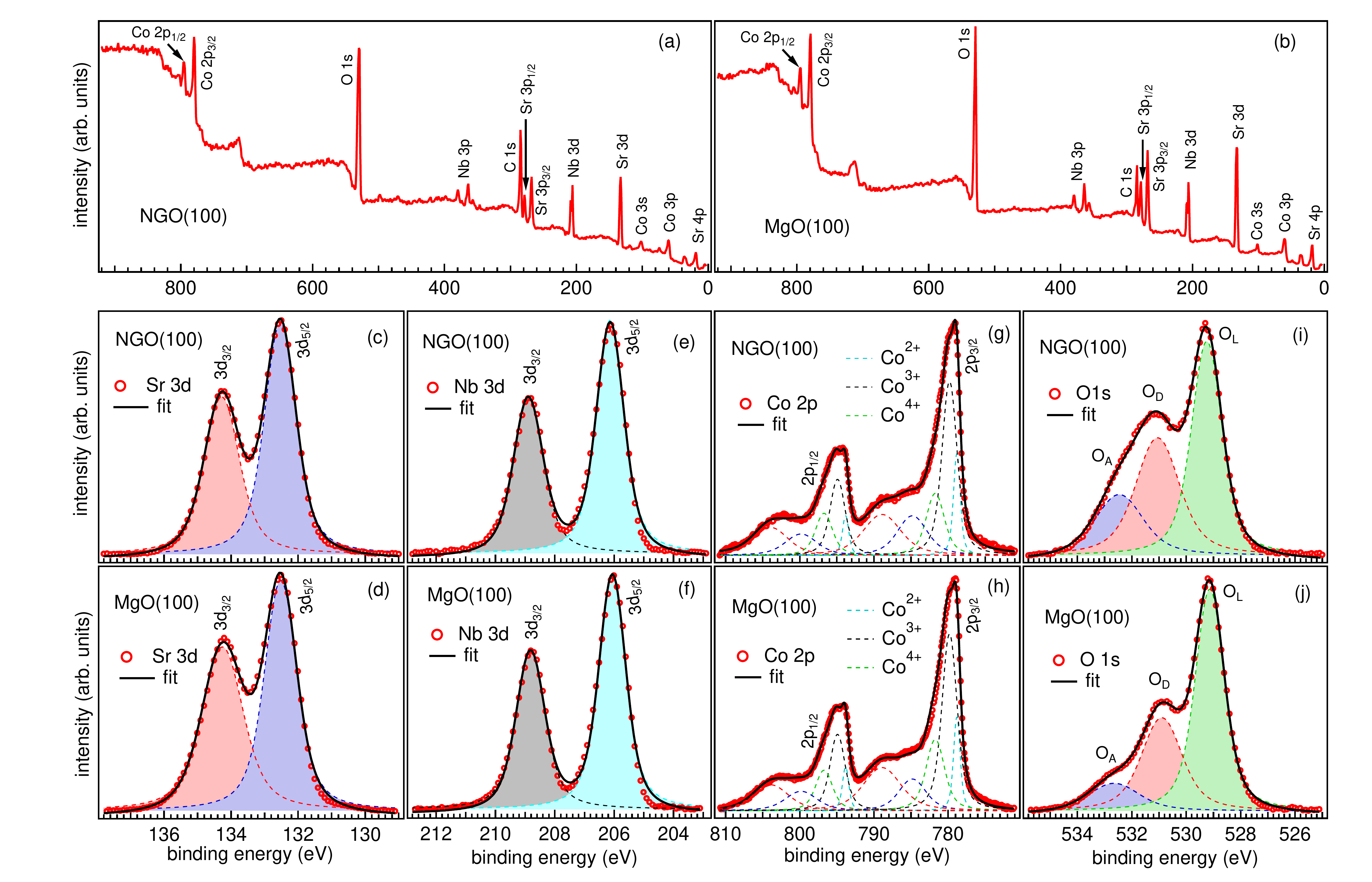}
\caption {(a, b) The XPS survey spectra, (c, d) Sr 3$d$ core-level, (e, f) Nb 3$d$ core-level, (g, h) Co 2$p$ core-level, and (i, j) and O 1$s$ core-level spectra of the Sr$_2$CoNbO$_6$ thin films grown on NGO(100) and MgO(100) substrates, respectively.} 
\label{Fig4_XPS}
\end{figure*}

To understand this observed change in the conduction mechanism, the x-ray photoelectron spectroscopy (XPS) measurements are performed on these samples \cite{Vasquez_JESRP_91, DhakaSS09}. Figs.~\ref{Fig4_XPS}(a, b) show the XPS survey spectra of Sr$_2$CoNbO$_6$ films grown on NGO(100) and MgO(100) substrates, respectively where all the prominent peaks in the spectra are assigned to the binding energies (BEs) of the constituent elements of the target material, discarding the possibility of any elemental impurity in both the samples. Figs.~\ref{Fig4_XPS}(c, d) represent the Sr 3$d$ core level spectra of these samples, which show a separation of 1.7~eV between 3$d_{3/2}$ and 3$d_{5/2}$ components resulting from the spin-orbit coupling. The peak position of these components (Sr 3$d_{5/2}$ at 132.5~eV  and Sr 3$d_{3/2}$ at 134.3~eV) indicate the presence of Sr in 2+ valence state in both the samples \cite{Vasquez_JESRP_91, Sosulnikov_JESRP_92}. Further, the recorded Nb 3$d$ core-level spectra are shown in Figs.~\ref{Fig4_XPS}(e) and (f) for films gown on NGO(100) and MgO(100) substrates, respectively. The comparison of the observed peak position of the spin-orbit splitted components (Nb 3$d_{5/2}$ at 206.1~eV  and Nb 3$d_{3/2}$ at 208.9~eV) with the reported values in the literature indicate the presence of the Nb predominantly in 4+ valence state for both the samples \cite{Bahl_JPCS_75, Wong_PRB_14}. The presence of Nb in tetravalent state indicate the strong possibility of the oxygen deficiency in the thin film samples as compared to the bulk Sr$_2$CoNbO$_6$, where our recent XAS measurements indicate the presence of Nb purely in the pentavalent state  \cite{Kumar_PRB3_22}. These oxygen deficiencies are expected to play a key role in governing the underlined transport properties of these samples, as discussed below.  

Furthermore, the Co 2$p$ core level XPS spectra are measured to understand the effect of the substrate induced strain and possible oxygen non-stoichiometry on the chemical environment of Co, as shown in Figs.~\ref{Fig4_XPS}(g) and (h) for the films grown on NGO(100) and MgO(100) substrates, respectively. We observe a clear splitting in the main peaks for both the 2$p_{1/2}$ and 2$p_{3/2}$ components, indicating the presence of the two different valence states of Co in both the samples. Further, the deconvolution of the spectra reveal the presence of an additional component, as evident from the asymmetrycity in the main peaks towards the higher binding energy in Figs.~\ref{Fig4_XPS}(g) and (h). The position of these three components are presented in the Table I, which resembles well with the reported values for Co$^{2+}$, Co$^{3+}$, and Co$^{4+}$ \cite{Biesinger_ASS_11, Chung_SS_76, Dupin_TSF_76}. This deviation in the valence states of Co in the thin films (from  the 3+ state in the bulk Sr$_2$CoNbO$_6$ \cite{Kumar_PRB3_22}) further indicate the possibility of strain induced oxygen non-stoichiometry in both the samples. Interestingly, the ratio of Co$^{2+}$ to Co$^{4+}$ is higher (0.61) in case of the film grown on NGO(100) substrate as compared to that (0.47) on MgO(100). An asymmetric in-plane compressive strain in case of NGO(100) substrate results in the elongation in the out-of-plane lattice  parameter of Sr$_2$CoNbO$_6$, which possibly stabilize the Co$^{2+}$ in the system due to its larger ionic radii as compared to Co$^{3+}$ \cite{Shannon_AC_76}. This small change in the fraction of the Co$^{2+}$ with respect to Co$^{4+}$ with the growth orientation may also result from the fitting procedure due to the over parameterization resulting from the several components present in these samples. However, the oxygen 1$s$ core-level XPS spectra discussed below clearly validate this point, which show the higher oxygen deficiencies in case of the film grown on the NGO(100) substrate as compared to MgO(100) substrate. Here, it is important to note that Co$^{2+}$ is insulating in nature due to the lowest possible oxidation state of Co, which significantly suppress the conduction channels in the sample \cite{Kumar_PRB1_20}. This results in the lower electronic conductivity of the film grown on the NGO(100) substrate as compared to that on MgO(100) substrate, as evident from the $\rho$-T measurements, presented in Fig. \ref{Fig2_rho}(a). Further, we observe the two broad satellite features around 784.8~eV and 788.9~eV which can be assigned to the Co$^{2+}$ and Co$^{3+}$ states, respectively \cite{Chung_SS_76}. Here, it is well known that Co$^{2+}$ shows much stronger satellite feature as compared to Co$^{3+}$; however, significantly large fraction of the Co$^{3+}$ as compared to Co$^{2+}$ in the present case results in the comparable strength of the satellite features for both the states. 

\begin{table}
\centering
		\label{tab:Oxygen}
		\caption{The fitting parameters of the Co $2p$ ($2p_{3/2}$) and O 1$s$ core-level spectra of Sr$_2$CoNbO$_6$ films deposited on the NGO(100) and MgO(100) substrates.}

\begin{tabular}{p{1.5cm}p{1.5cm}p{1.5cm}p{1.5cm}p{1.5cm}}
		\hline
\hline
	 Substrate	& Peak &  Position & FWHM & Area \\
   (100) & & (eV) & (eV) & \\
\hline
& & Co $2p$ &\\
 NGO & 2+ & 778.8 & 1.1& 0.62 \\
 & 3+  & 779.8 & 2.4 & 2.36 \\
 & 4+ & 781.7 & 2.9 & 1.01 \\
 MgO & 2+ & 778.8 & 1.1 & 0.56 \\
 & 3+  & 779.8 & 2.4 & 2.40 \\
 & 4+ & 781.7 & 2.9 & 1.18 \\
\hline
& & O $1s$ &\\
 NGO & O$_L$ & 529.2& 1.3&1.61 \\
 & O$_D$  & 531.0& 1.9&1.26 \\
 & O$_A$ & 532.5& 2.1&0.71 \\
 MgO & O$_L$ & 529.1& 1.3&1.58 \\
 & O$_D$  & 530.9& 1.8&0.93 \\
 & O$_A$ & 532.6& 2.2&0.34 \\
\hline
\hline
\end{tabular}
\end{table}

\begin{figure*}
\includegraphics[width=1.03\textwidth]{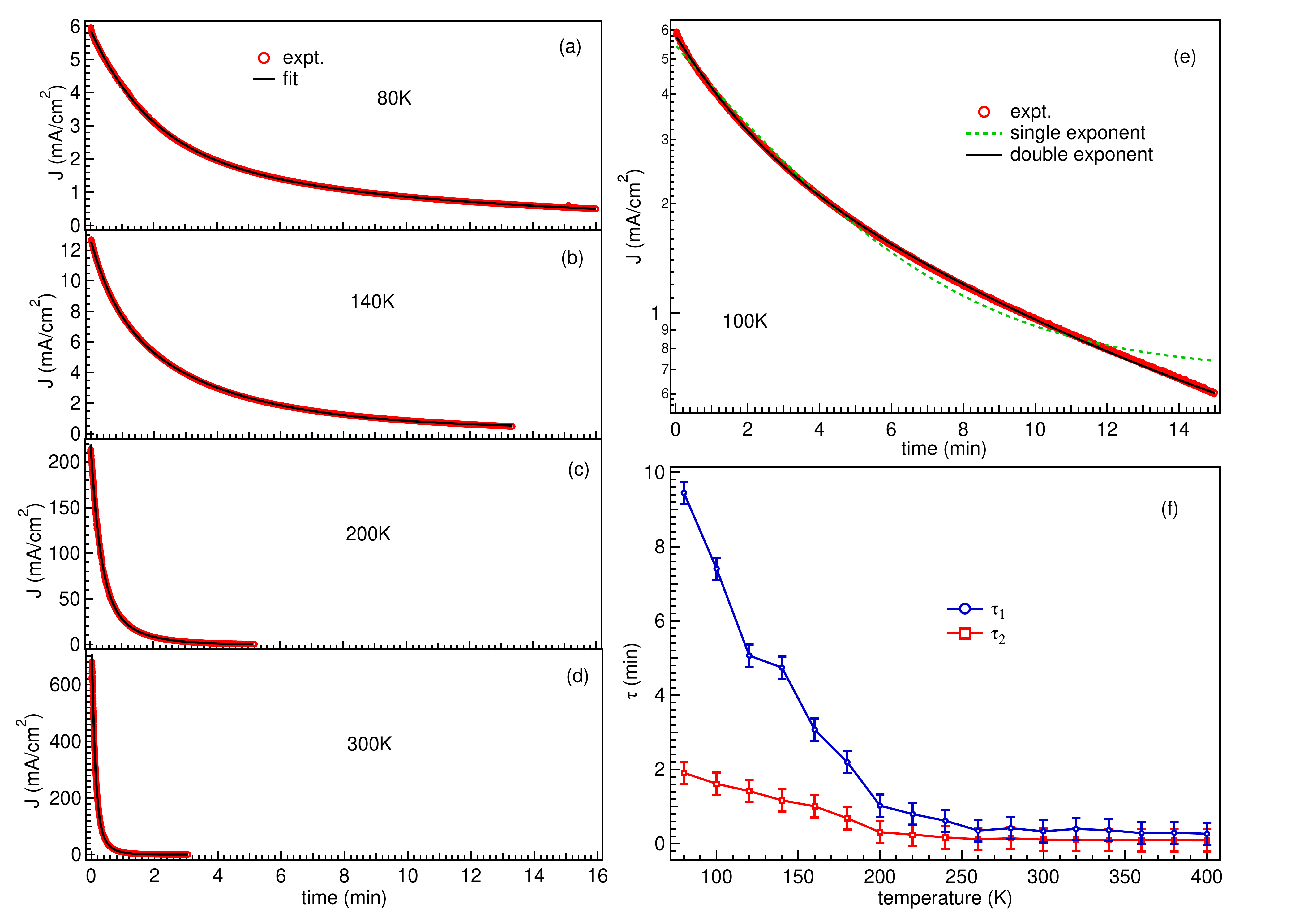}
\caption {(a--d) The time dependence of the current density after applying a step voltage of 500~V/cm for 3 minutes on the Sr$_2$CoNbO$_6$/NGO(100) sample at the selected temperatures. (e) Fitting of the $J-t$ curve at 100~K using single (green dashed curve) and double (solid black line) exponent models (see text for more details). (f) Temperature evolution of the relaxation times for both the relaxation processes.} 
\label{Leakage}
\end{figure*}

Moreover, we measured the O 1$s$ core-level XPS spectra to estimate the possible oxygen defects in these samples, as presented in Figs.~\ref{Fig4_XPS}(i, j). We find three well-resolved components for both the samples; however, two features at the higher BE are more clearly distinguishable in the case of film grown on MgO(100) substrate. Here, the first component around 529.1(1)~eV (O$_L$) is attributed to the lattice oxygen in the Co/Nb--O octahedra, whereas third component around 532.6(2) (O$_A$) is resulting from the chemisorbed oxygen atoms, i.e., surface contamination by organic molecules \cite{ShuklaPRB22}. Importantly, the central component around 531.0(1) (O$_D$) results either from the oxygen atoms with the formal charge less than -2$e$ due to the covalent character of the Co/Nb--O bonds \cite {Dupin_PCCP_2000, Wu_AIP_15} or presence of oxygen deficiencies in the samples \cite{Cui_PRB_19, Cho_PRM_19, Lei_JACS_14}. The intensity of this central component with respect to the lattice oxygen is much higher in case of NGO(100) film as compared to the MgO(100) [see Figs.~\ref{Fig4_XPS}(i, j)], which clearly indicate the higher oxygen deficiencies in the former. The integrated area ratio of O$_D$ and O$_L$, which can be used as the measure of the oxygen deficiencies in the thin film samples \cite{Cui_PRB_19, Cho_PRM_19}, is found to be 0.78 in case of NGO(100) substrate and 0.59 for the film grown on MgO(100) substrate. This higher oxygen deficiency in the film grown on NGO(100) substrate results in the larger concentration of Co$^{2+}$ ions and hence lower electronic conductivity as compared to the film grown on MgO(100) substrate, which is also evident from the Co 2$p$ core-level spectra, as discussed above. However, the exact quantification of oxygen vacancies using the high end techniques such as aberration corrected transmission electron microscopy (TEM) and/or positron annihilation can give much deeper insight into their role in governing the conduction mechanism of these samples \cite{Gauquelin_UM_17, Reiner_APL_15}.  

Interestingly, the colossal dielectric properties observed in Sr$_2$CoNbO$_6$ \cite {Wang_AIP_13} indicate its possible use in the energy storage devices, where the study of the leakage current is important to find its practical usefulness \cite{Nagaraj_PRB_99}. Thus, in order to estimate that, we apply a step voltage of 500~V/cm on Sr$_2$CoNbO$_6$/NGO(100) sample for 3 minutes and then record the current response as a function of time (until current drops to 1~nA), as shown in Figs.~\ref{Leakage}(a--d) at some representative temperatures. Interestingly, the current persists up to several minutes at the lower temperatures and decay more rapidly with increase in the temperature. This non-linear response of current is fitted with the sum exponent model defined as \cite{Podgorny_AIP_16}
\begin{eqnarray}
J(t)= \sum_{i=1}^{n}J_{mi}e^{-t/\tau_i}+J_0, 
\end{eqnarray}
where $J_{mi}$, $\tau_i$, and $J_0$ represent the initial current density, relaxation time, and steady-state current density, respectively and the summation is over the different relaxation processes. First, we try to fit the $J-t$ curves using the single exponent model. However, a significant deviation from the experimental data has been observed, as shown by the green dashed line in Fig.~\ref{Leakage}(e) for 100~K. Thus, two exponent mode with different relaxation times is used for the fitting and a nice agreement between the experimental and fitted curves can be seen, as represented by the solid black line in Fig.~\ref{Leakage}(e). This indicates the presence of two different relaxation mechanisms in the sample. Therefore, all the $J$-t curves at different temperatures are fitted using the two exponent model and the temperature evolution of the two relaxations times is presented in Fig.~\ref{Leakage}(f). It is interesting to note that the relaxation time $\tau_1$ is 5-6 times higher than $\tau_2$, which indicate significantly different origin of the two processes. The similar behavior of the $J-t$ curves with two relaxation processes has been also observed in the polycrystalline films of Pb(Zr$_{0.48}$Ti$_{0.52}$)O$_3$, which are speculated to originate from the bulk and grain boundaries/sample-electrode interface \cite{Podgorny_FE_12, Podgorny_AIP_16}. The effect of grain boundaries is expected to be negligible on the single crystalline film and the two relaxation processes are most likely originate from the sample and interfacial polarization in the present case. However, further investigations of the defects, twinning in the film, oxygen vacancies, carrier hopping, traps filling, etc., can shed light on the nature of these relaxation processes \cite{Stolichnov_JAP_98, Yoon_JAP_2000, Pintilie_PRB_07, Rojac_PRB_16}. Moreover, both the relaxation times decreases with increase in the temperature up to around 200~K and then attain a very small value ($<$ 1 min) at the higher temperatures. This is due to the enhancement in the electronic conductivity and hence availability of the more charge carriers at the higher temperature. Thus, a large dielectric constant observed in Sr$_2$CoNbO$_6$ near the room temperature \cite{Wang_AIP_13} is possible accompanied with the higher electronic conductivity in the sample, which limit its practical use in the charge storage applications and hence further engineering of the electronic band gap is required for its practical use. 

\begin{figure}[h] 
\includegraphics[width=3.35in]{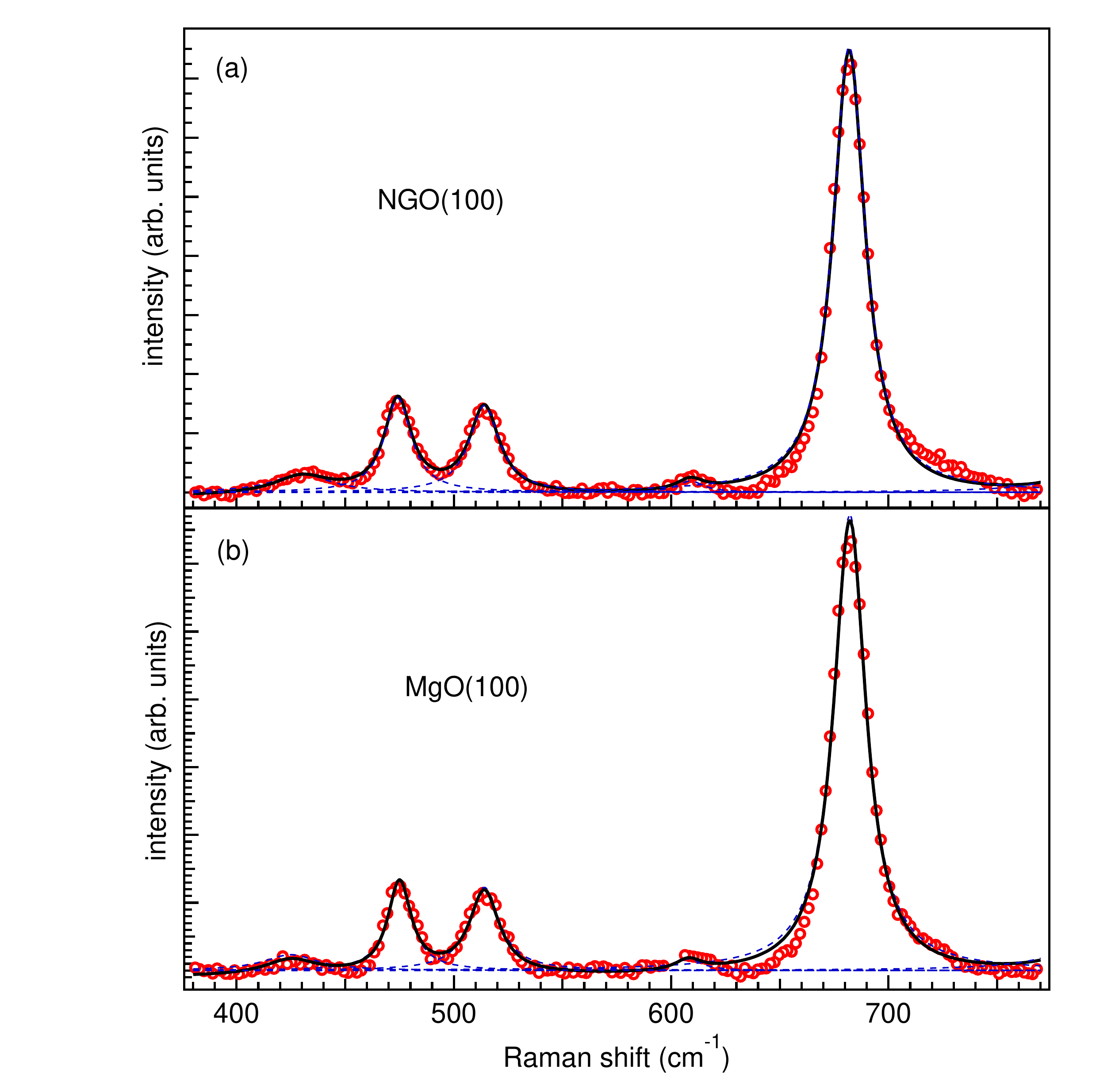}
\caption {The room temperature Raman spectra of Sr$_2$CoNbO$_6$ thin films grown on (a) NGO(100) and (b) MgO(100) substrates using the 514.5~nm excitation wavelength.}
\label{Fig6_Raman}
\end{figure}

Finally, we show the Raman spectra for the films grown on NGO(100) and MgO(100) substrates, in Figs.~\ref{Fig6_Raman}(a, b), respectively. We observe several Raman active modes between 380--770cm$^{-1}$ for both the samples, unlike bulk Sr$_2$CoNbO$_6$ which shows only two very weak Raman modes \cite{Kumar_PRB1_20}. This suggest the reduction in the crystal symmetry in the thin film samples as compared to the bulk Sr$_2$CoNbO$_6$ possibly due to the distortion in the (Co/Nb)O$_6$ octahedra resulting from the substrate induced strain. For example, the the presence of the three Raman modes between 400-550 cm$^{-1}$ represent the oxygen bending modes in the $P2_1/n$ (monoclinic) symmetry, as no Raman active modes are expected at this wave number for the $I4/m$ (tetragonal) space group of the bulk Sr$_2$CoNbO$_6$ material \cite{Andrews_Dalton_15, Kumar_PRB1_20, Kumar_JAP_20}. Note that the lowering in the crystal symmetry in the double perovskite oxides is usually accompanied with the enhancement in the B-site ordering. Thus, the presence of these Raman active modes suggests the growth of B-site ordered thin films from the almost disordered bulk Sr$_2$CoNbO$_6$ sample \cite{Kumar_PRB1_20, Kumar_PRB3_22}. This is consistent as the substrate induced strain is considered an effective way to grow the ordered thin films from the disordered target materials \cite{Chakraverty_APL_13, Chakraverty_PRB_11, Yoshimatsu_PRB_15, Kleibeuker_NPGAM_17}. However, no significant change is observed in the Raman spectra for both the sample, which indicate that any change in the degree of the octahedral distortion and/or B-site ordering due to change in the growth orientation can not be probed using the unpolarized Raman spectroscopy in the present case and the polarization dependent Raman spectra can be more useful \cite{Iliev_PRB_07}.  

\section{\noindent ~Conclusions}

The oxygen stoichiometry and hence the resulting electronic and transport properties of Sr$_2$CoNbO$_6$ have been engineered with change in the growth orientation of the epitaxial thin films using pulsed laser deposition. The films of Sr$_2$CoNbO$_6$ have been grown along $a$ and $c$-axis on orthorhombic NGO(100) and cubic MgO(100) substrates, resulting in the asymmetric compressive and symmetric tensile in-plane strain, respectively. The film grown on the NGO(100) substrate have the higher degree of the oxygen deficiency, which results in the larger fraction of Co in 2+ valence state and hence derive the system toward the insulating regime as compared to that grown on MgO(100) substrate. The XPS measurements shows the presence of Co in the 2+, 3+ as well as 4+ valence states, resulting from the in-plane compressive (tensile) and hence out-of-plane tensile (compressive) strain in case of the film grown on the NGO(100) [MgO(100)] substrate. Moreover, the divalent and tetravalent states of Sr and Nb, respectively, are found be remain invariant with the growth orientation in the two cases. Interestingly, the investigation of the leakage current indicate that the colossal dielectric properties observed in bulk Sr$_2$CoNbO$_6$ near room temperature originate from the higher electronic conductivity in the sample. Moreover, the Raman spectra evident the significant reduction in the crystal symmetry in the thin films as compared to the bulk Sr$_2$CoNbO$_6$. 

\section{Acknowledgment}

AK acknowledges the UGC India for fellowship and the physics department of IIT Delhi for providing the the XRD, AFM, and PPMS EVERCOOL facilities, and central research facilities (CRF) of IIT Delhi for providing the Raman spectrometer. AK thanks Rishabh Shukla for his help during the thin film deposition. AK also thanks Ploybussara Gomasang for help in the XPS measurements at the Shibaura Institute of Technology, Japan, which were supported by Sakura Science Program (aPBL). RM thanks IUAC for providing the experimental facilities. The PLD instrument used for thin film growth is financially supported by IIT Delhi through seed grant with Reference No. BPHY2368 and SERB-DST through early career research (ECR) award with Project Reference No. ECR/2015/000159. RSD also acknowledges the SERB–DST for financial support through a core research grant (project reference no. CRG/2020/003436).



\end{document}